| | |
|---|---|
| **Title** | **Self-organized patterns by a DC pin liquid anode discharge in ambient air: Effect of liquid types on formation** |
| **Authors** | Shiqiang Zhang, and Thierry Dufour |
| **Affiliations** | LPP, CNRS, UPMC Univ. Paris 06, Ecole Polytech., Univ. Paris-Sud, Observatoire de Paris, Université Paris-Saclay, Sorbonne Universités, PSL Research University, 4 Place Jussieu, 75252 Paris, France. E-mail: thierry.dufour@sorbonne-universite.fr |
| **Ref.** | Physics of Plasmas 25, 073502 (2018) |
| **DOI** | https://doi.org/10.1063/1.5030099 |
| **Abstract** | A pin liquid anode DC discharge is generated in open air without any additional gas feeding to form self-organized patterns (SOPs) on various liquid interfaces. Axially resolved emission spectra of the whole discharge reveal that the self-organized patterns are formed below a dark region and are visible mainly due to the $N_2$ ($C^3\Pi$-$B^3\Pi$) transitions. The high energy $N_2$ (C) level is mainly excited by the impact of electrons heated by the local increased electric field at the interface. For the first time, the effect of the liquid type on the SOP formation is presented. With almost the same other discharge conditions, the formed SOPs are significantly different from HCl and $H_2SO_4$ liquid anodes. The SOP difference is repeated when the discharge current and gap distance change for both liquid anodes. The variations of SOP size and discretization as a function of discharge current and gap distance are discussed and confirm that different SOPs are formed by the HCl liquid anode from tap water or the $H_2SO_4$ liquid anode. A possible explanation is brought up to explain the dependence of SOPs on the liquid type. |

# I. Introduction

To date, plasma liquid interactions (deionized water, tap water, saline, or electrolyte solution) draw increasing attention due to their great potential applications in plasma medicine,[1,2] plasma agriculture,[3,4] nanomaterials synthesis,[5,6] and water purification.[7] In plasma activated media (PAM), longlived reactive species such as $H_2O_2$, $HNO_2$, $HNO_3$ and $ONOO^-$ are generated or diffused into liquid from the plasma. As an example, the production of $H_2O_2$ in an aqueous medium can result from a gas discharge plasma interaction as found in Ref. 8. $HNO_2$ is another paramount species verified to diffuse from the plasma to the liquid via the interface region.[9] As a result, it is of utmost importance to investigate the physical and/or chemical processes at the interface that determines the aqueous amounts of species in liquid.[2]

| Type of power supply | SOPs formed on a … | |
|---|---|---|
| | Plasma-solid interface | Plasma-liquid interface |
| **AC** | • Enclosed DBD[13-16]<br>• APPJ impinging on ITO in open air[17] | • Pin liquid electrode discharge in open air[18, 19] |
| **DC** | • CBLD in noble gas on the cathode[20-24]<br>• DBD with a semiconductor electrode[25]<br>• Glow discharge at at 1 atm on the anode[26,27] and low pressure[28-32] | • Pin liquid electrode discharge in the air with auxiliary gas[33-35] and without[36-38]. |

*Table 1. Types of plasma devices with formed self-organized patterns.*

Plasma self-organized patterns (SOPs) among physical phenomena are likely to appear on a liquid or a solid interface, considering AC or DC power supplies as shown in Table I. It should be noted that besides the SOP formed at the plasma-liquid (or even solid) interface, patterns can be formed in the plasma as seen from the lateral side of the discharge chords.[10–12]

In parallel to experimental works, SOPs have been the subject of several theoretical studies considering various discharge configurations such as dielectric barrier discharge (DBD) discharge,[39,40] arc discharge,[41,42] and DC glow discharge.[43,44] Benilov modeled SOPs in DC glow discharge and/or arc discharge based on the first-principles equations containing all the relevant physical mechanisms and proposed that the cathode sheath instability is responsible for SOP formation on the cathode.[43] Bieniek et al. also applied the same approach to simulate SOPs formed on the anode of DC glow discharge and found that the anode SOPs are related to the change in the sign of the near anode voltage.[44] Trelles obtained SOP results, which is in quite good agreement with experimental observations but for arc discharge.[45] Raizer and Mokrov owed SOP formation to the instabilities caused by thermal expansion and/or electric field redistribution.[15] Gopalakrishnan et al. studied solvated electrons but not SOPs in a pin liquid discharge in argon by a particle-in-cell/Monte Carlo model.[46] To date, no simulation work reports SOP formation in pin liquid discharge in air to the best of the authors' knowledge.

Obviously, the SOPs on the cold solid anode of DC discharges are the closest relative of SOP on cold liquid anodes of DC discharges that are studied in this work. However, there are still differences between them. Replacing a metal electrode with a liquid electrode in the vicinity of the air discharge increases its water content,[47] which changes the stability of the discharge48 and hence SOP formation experimentally. Additionally, due to larger specific heat capacity of water than the metals (generally one order of magnitude larger), the gas temperature decreases, making the discharge not easy to turn in strong filamentary or even arc discharge.[49] The auxiliary gas, such as helium,34 could also lower the gas temperature and establish the comparatively stable discharge and reproducible SOP formation.

Comparing the work of Refs. [36] and [37] to that of Ref. 34, for a pin liquid electrode discharge configuration, the use of auxiliary gas is not necessary to form the luminous SOPs in open air. Moreover, electric conductivity (EC) is experimentally verified to determine the SOP formation with a NaCl solution electrode in Ref. 34 and a water electrode in Ref. 37. However, the effect of the liquid type on SOP formation is never reported yet. Considering the luminescence, although the spectra of SOP under different discharge setups are observed, there is no underlying mechanism for the analysis of the observed spectra either for the moment.

In this work, the dependence of SOPs on liquid types (different chemical compositions of anode liquid) is reported for the first time. Using the specific pin liquid anode discharge, the line-of-sight integrated emission spectra with axial resolution are presented, and the underlying mechanism of the SOP spectra is analyzed. Moreover, the behavior of the self-organized patterns with discharge current and the gap distance under different liquid anodes (tap water, HCl solution, and $H_2SO_4$) confirms the dependence of the SOPs on the liquid type. Finally, an attempting explanation is brought on the liquid type influence on SOP formation.

# II. Experimental setup

The discharge setup is shown in Fig. 1. A DC power supply (FuG, type HCN 350M-6500, 0–6.5 kV, 0–30 mA) is connected to a high power ballast resistor (Sfernice, RSS 30 265, 111 kΩ) to stabilize the atmospheric pressure plasma. An aluminum pin is employed as a cathode above the liquid anode with varying gap distances. The pin has a diameter of 3.2mm and is sharpened at the end. Stainless steel submerged in the liquid (electrolyte solution), which is filled in a quartz vessel (height, 13 mm; diameter, 57 mm), is connected to the anode of the power supply. A spectrometer (Andor Shamrock SR 303i-A) is used to determine the line-of-sight integrated plasma emission from the lateral side of the discharge chord. The electrical conductivity (EC) is quantified using a conductivity meter (HANNA HI-87314).

The chemical products used in this work are H2SO4 acid (500 ml, for the determination of nitrogen, >97.5%, CAS No.: 7664-93-9; Sigma Aldrich/Merck), HCl acid (1 L; MIEUXA), and HF acid (for analysis, EMSUR, 48%; Sigma Aldrich/Merck).

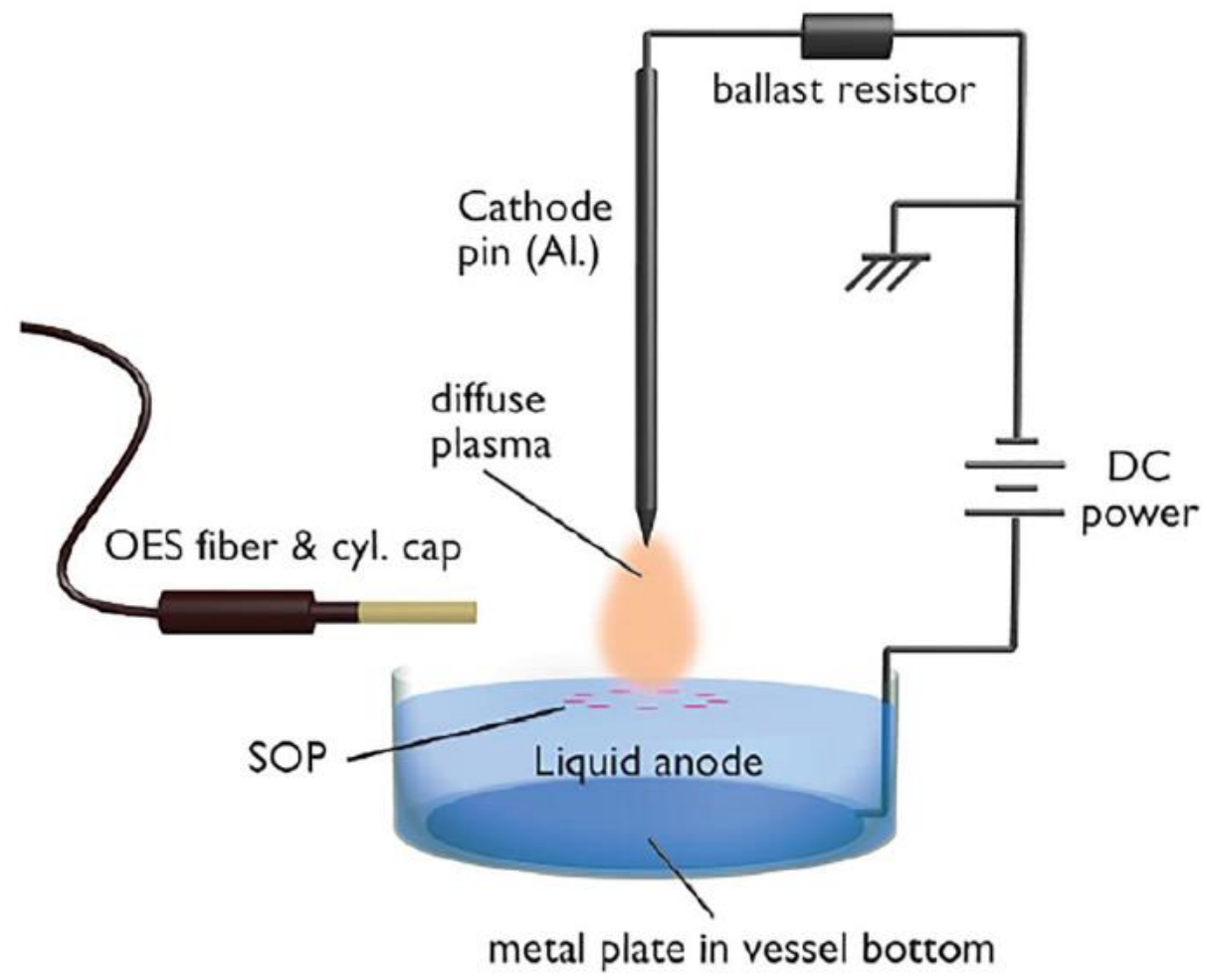

***Fig. 1. Schematics of the pin liquid anode DC discharge setup.***

# III. Results & Discussions

## III. A. Axially resolved optical emission

### III.A.1. OES spectra results

A typical DC pin liquid anode discharge is pictured in Fig. 2(a). It is clearly observed that a pear-shaped and homogeneous discharge is emitted by the cathode pin and a self-organized pattern (SOP) of plasma lying on the liquid anode interface. Also, a dark region separates these two emissive plasma regions, with a typical thickness of 1mm at the center and 2mm in periphery. Finally, the apparent color of the diffuse plasma is rather yellowish, while it is purplish for the SOP. In order to perform a finer analysis, the discharge is characterized by optical emission spectroscopy. The optical fiber (capped with the 1mm inner diameter capillary) is placed at 6 different vertical positions between the anode and the cathode, while the interelectrode distance remains fixed at 6mm upon the entire experiment. The discharge emission is estimated between 200 and 700 nm every mm along the vertical axis and reported in the 200–500 nm range in Fig. 2(b). The main bands identified are the $N_2$ second positive system (SPS, $C^3\Pi$-$B^3\Pi$), the OH ($A^2\Sigma^+$-$X^2\Pi$) transition, and the NO ($A^2\Sigma^+$-$X^2\Pi$) system. Figure 3 reports the axial profile of these bands along gap distances and the total plasma emission in the 200–700nm range. It is worth mentioning that no hydrogen (656 nm), oxygen (777 nm, 844 nm), or even nitrogen atomic lines (>700 nm) could be detected in any optical fiber position with this aluminum pin electrode.

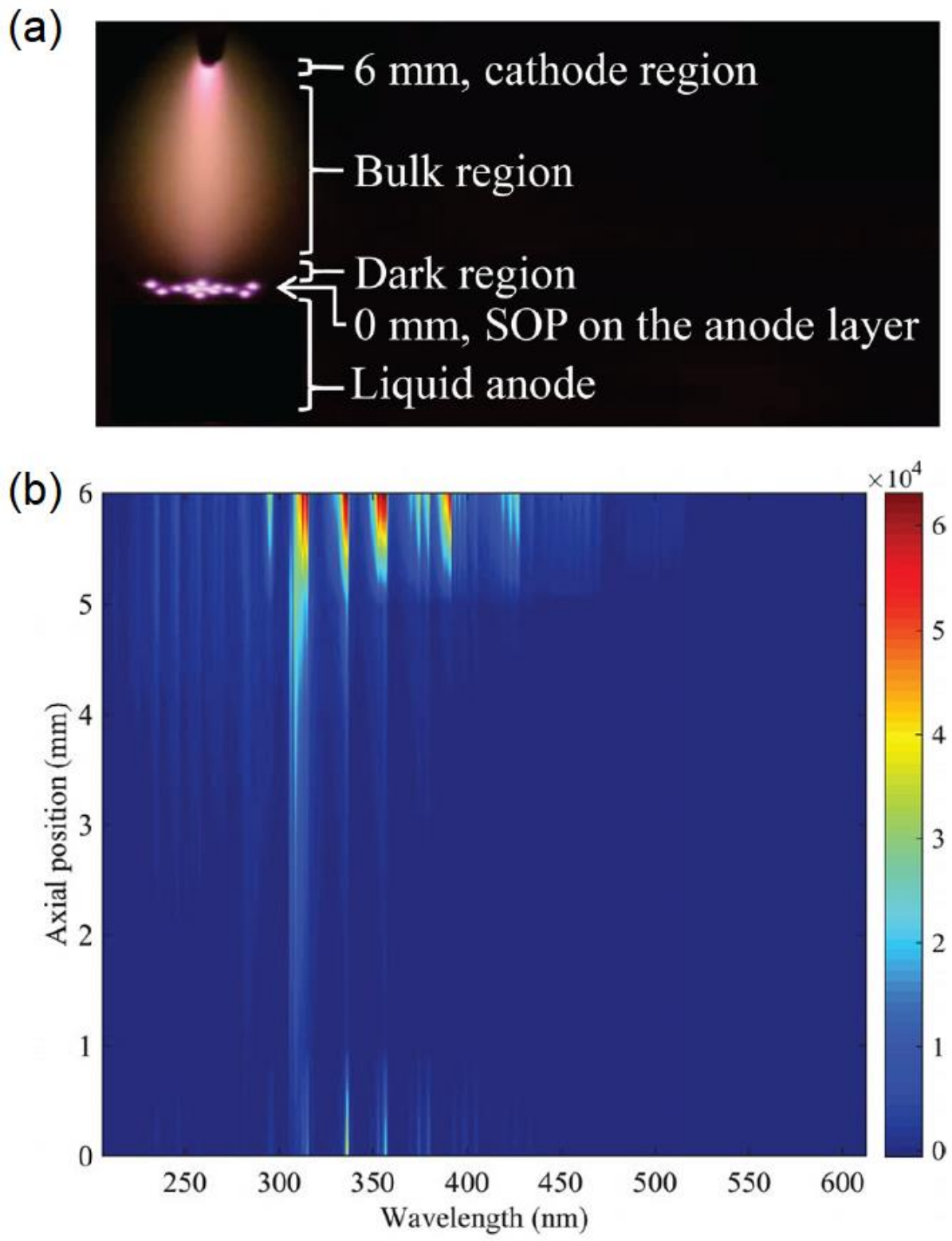

*Fig. 2. (a) Picture of the pin liquid anode discharge taken using a camera and (b) the axially resolved optical emission spectrum of the discharge. The exposure time for (b) is 50 ms, and the employed grating is 600 grooves/mm. The discharge voltage is 1150 V, the current 18.5 mA, and EC=2.66 mS/cm. The emission spectra are acquired from axial 0mm to 6mm with a 1mm step.*

### III.A.2. Analysis of pin liquid discharge and SOP emissions

Combining camera pictures and optical emission spectroscopy (OES) spectra allows us to state that the discharge is stratified into 4 main regions: (i) between 5 and 6 mm, a cathode region which is highly emissive with several molecular nitrogen and nitric oxide bands, as well as the OH band headed at 310 nm; (ii) between 5 and 1 mm, a bulk region characterized by an overall decay of the discharge emission, in particular the $N_2$ SPS bands whose emissions are wallowed in the background. The OH band emission decreases as one moves closer to the liquid anode but much slower, so that it is still significant at 1mm compared with $N_2$ SPS bands; (iii) between roughly 1 and 0.5mm appears the dark region: its thickness is much thinner along the main vertical axis than on the lateral edges of the discharge. This region is correlated not only with the minimum emission of $N_2(C^3\Pi\text{-}B^3\Pi)$ and NO $(A^2\Sigma^+\text{-}X^2\Pi)$ transitions, as shown in Fig. 2(b), but also with the total plasma emission, which is the lowest at 1mm (Fig. 3); (iv) finally, between 0.5mm and 0 mm, the anode region (or layer) is characterized by an unexpected significant rise of $N_2(C^3\Pi\text{-}B^3\Pi)$, while OH $(A^2\Sigma^+\text{-}X^2\Pi)$ and NO$(A^2\Sigma^+\text{-}X^2\Pi)$ are negligible (1 order of magnitude smaller). SOPs are generated in this anode region and seem to mostly result from physicochemical pathways, inducing the excitation of molecular nitrogen. The fact that $N_2$ $(C^3\Pi\text{-}B^3\Pi)$ dominates the emission of SOPs on the liquid interface is consistent with other different types of discharge setups, such as a similar type in Refs. 36 and 50, a plasma jet colliding with the water surface,[51] surface discharge over water in Ref. 52, water falling film DBD in Ref. 53, or in other types of pin liquid discharge systems.[19,34,54]

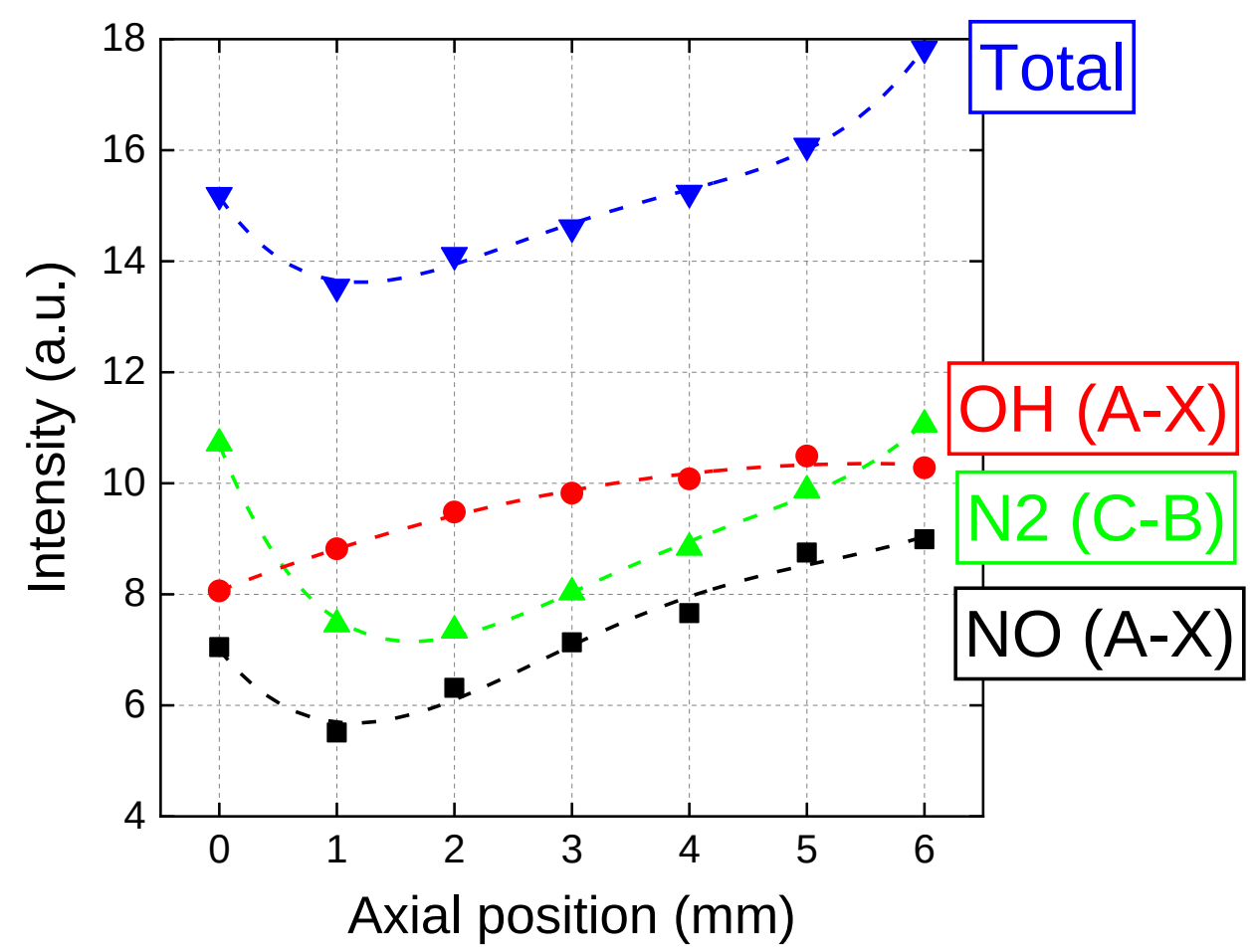

***Fig. 3. Axial distribution of OH(A-X) (309 nm), NO(A-X)(236.3 nm), $N_2(C^3\Pi\text{-}B^3\Pi)$ (337.1 nm), and total emission with the intensity on a logarithmic scale.***

Note that in a similar DC pin liquid anode discharge operating also in open air,[50] the spatial stratification is different. Indeed, the authors obtained a comparatively thicker luminous anode region and no evident dark region above the luminous anode region. In addition, they have Faraday dark space and a cathode space which do not appear in this work. The disparity might be due to the material of the pin electrode, where an aluminum pin is used here rather than stainless steel in their work. As a matter of fact, the authors also tried to change the polarity of the discharge, to apply a copper pin or a carbon pin, or change liquid to a metal plate as above. However, when the polarity changed, there are no such SOP patterns formed with the aluminum (Al) pin anode and liquid cathode, only to find a strong discharge column. When the copper or carbon pin cathode is used, the formed SOP on the liquid anode is not stable enough even to perform the measurements. The used configuration is the most stable type we tried. For the structure of the pin cathode and liquid anode, the authors presume that the pin material and fine shape of the pin tip greatly affect stability and SOP formation.

The plausible chemical pathways producing OH and NO radicals and excited $N_2$ molecules are listed in Table II.

| | Reactions | Typical rate constants | References |
|---|---|---|---|
| $R_1$ | $e^- + H_2O \rightarrow OH(A/X) + H + e^-$ | $10^{-18}$–$10^{-16}$ $m^3 s^{-1}$ [a] | 55 and 56 |
| $R_2$ | $e^- + OH(X) \rightarrow OH(A) + e^-$ | $10^{-14}$–$10^{-13}$ $m^3 s^{-1}$ [a] | 56 and 57 |
| $R_3$ | $e^- + H_3O^+ \rightarrow OH + H_2$ | $10^{-13}$ $m^3 s^{-1}$ | 58 |
| $R_4$ | $e^- + O_2 + M \rightarrow O_2^- + M[O_2, N_2]$ | $10^{-43}$–$10^{-42}$ $m^6 s^{-1}$ | 59 |
| $R_5$ | $e^- + N_2(X) \rightarrow N_2(C) + e^-$ | $10^{-19}$–$10^{-16}$ $m^3 s^{-1}$ [a] | 56 and 60 |
| $R_6$ | $N_2(A) + N_2(A) \rightarrow N_2(C) + N_2(X)$ | $10^{-16}$ $m^3 s^{-1}$ | 61 |
| $R_7$ | $N_2(A) + H_2O \rightarrow N_2 + H + OH$ | $\leq 10^{-20}$ $m^3 s^{-1}$ | 62 |

***Table II. Main chemical reactions in overall discharge and SOP region. Rate constants for electron impact dissociation and excitations are calculated from the cross-section in the given references with electron temperature $T_e$=1-3 eV.***

**a. Overall.** In an air discharge, OH generation by thermal dissociation of water molecules could be negligible, as the rate coefficient is somewhat 4 orders lower than that of the electron dissociation ($R_1$ and $R_3$ in Table II). This is verified even for the gas temperature around 2000 K.[63]

**b. Cathode region (5–6 mm)**. In the cathode region, the electrons are heated in the thin cathode fall layer (for the aluminum cathode in air, $V_n \approx 300V$, $p \approx 3$ µm, and $E \approx 10^8$ V/m). Here, high energetic electrons can dissociate water vapor from the surrounding atmosphere to produce most of the OH radical ($R_1$, ≈5 eV). This pathway may not occur further in the bulk region where the electron temperature is estimated to be less than 2 eV.[64] Another source of OH radical is from electron dissociative recombination ($R_3$), as hydrated $H_3O^+$ readily pops out in such a pin liquid anode device.[65] Note that the generation of OH by $N_2$(A) dissociation with $H_2O$ ($R_7$) in ambient air is only comparable with the generation via $R_1$ or $R_2$ if the $N_2$(A) density is larger than $10^{19}$ $m^{-3}$.[66]

**c. Bulk region (1–5 mm).** As shown in Fig. 2(b), this region is solely dominated by the emission of OH ($A^2\Sigma^+$-$X^2\Pi$). This transition results from (i) the diffusion of the OH radicals produced in the cathode region and (ii) the excitation of the ground state OH to OH(A) by the electron impact ($R_2$).[48] As it approaches the liquid anode ($T_{aq} \approx 330K$), it is not surprising that the gas temperature decreases and the electric field intensity maintains a comparatively low level in the bulk region (the average electric field is $1.5 \times 10^5$ V/m considering the cathode fall). The reduced gas temperature and the electron temperature would enhance the three-body electron attachment to oxygen ($R_4$, the rate could be $10^{-17}$ $m^3.s^{-1}$, provided the air number density of $10^{25}$ $m^{-3}$). A decrease in the number of free electrons reduced the possibility of excitation of the ground OH radical, which makes the plasma emission from OH(A) decreases when it is further away from the pin electrode. This also leads to a decrease in the emission intensity of reactive nitrogen species [$N_2$(C) or NO(A)] when it approaches from the pin electrode to the anode.

**d. Dark region (0.5–1mm).** The formation of the dark region around the 1mm axial position correlates with a larger extent with the minimum emission of $N_2$ ($C^3\Pi$-$B^3\Pi$), as shown in Fig. 3. At atmospheric pressure, $N_2$ (C) is mainly excited by the direct electron impact of the ground state $N_2$ (X) ($R_5$), such as in a RF atmospheric pressure plasma jet (APPJ) effluent in open air with Te≈1.5 eV,[67] as opposed to the pooling reaction ($R_6$).[56,68] So, the decrease in free electrons also leads to reduced emission of $N_2$ ($C^3\Pi$-B3$\Pi$). The minimum emission of the dark region is found to be consistent with the results of an AC plate liquid electrode discharge in open air.[69]

**e. Anode region (luminous anode layer, 0.5–0 mm).** The anode region is marked by a sudden significant increase in the optical emission and the parallel formation of SOP [Figs. 2(b) and 3]. At the interface, the $N_2$ ($C^3\Pi$-$B^3\Pi$) emission mainly leads to the visible self-organized patterns [Fig. 2(b) at 0 mm]. Figure 3 shows that the intensity of $N_2$ SPS transition is significantly larger than the emission of the plasma bulk zone [mainly from OH(A)], hence, explaining why the so-called dark region is formed.

If one compares direct impact reactions to produce $N_2$(C) via $R_5$ or OH(A) via $R_2$, it turns out that the electron energy is somewhat two times larger to produce $N_2$(C) [≈11 eV (Ref. 70)] than to produce OH(A) [≈4 eV (Ref. 71)]. Therefore, more energetic electrons arise in the anode region, hence enabling the excitation of $N_2$(C) at the interface. At collision-frequent atmospheric pressure, the ohmic heating is the dominant heating mechanism of electrons,[72] which means that energetic electrons are heated by a local increased electric field. Nevertheless, no investigation of the electric field at the interface under such discharge is reported yet to verify the local increased electric field, although in Ref. 46 the model work on a DC pin liquid anode discharge in argon atmospheric pressure presented that at the interface, the electric field increased from $\approx 1.5 \times 10^5$ V/m to $\approx 6.5 \times 10^5$ V/m on the plasma side on a 0.1mm scale. Further measurement on the electric field is required.

Even if the spectrum of SOPs is characterized using OES, the mechanism for the regularity of SOP is not clear. The regularity of SOPs might also be the type of diffraction pattern of ionization waves40 launched from the pin. As ionization waves impinge at the liquid surface, the electric field at the SOPs is enhanced due to the interference of ionization waves and electrons are heated, which excites $N_2$(C) to the $N_2$(B) state as described above. However, this hypothesis needs further validation. In Sec. III B and III C, SOP behavior is studied by varying the discharge current and gap distance to better understand the SOP formation mechanism and to benchmark the SOP liquid type dependence.

## III. B. Effects of the discharge current on SOP

### III.B.1. Voltage-current characteristics and SOP appearance

The V–I characteristic of the DC discharge is shown in Fig. 4(a) where the discharge current is decreased from 37mA to 6mA by reducing the applied voltage. Decreasing the current induces a rise in the voltage up to a critical value of 1230V at 26mA. Then, a sudden voltage drop (or jump) is observed at 26mA immediately followed by a second voltage increase with a maximum as high as 1420V at 6mA. The sudden voltage drop at 26mA delimits the two regions shown in Fig. 4(a). In parallel to the V–I characteristic, pictures of the SOPs have been taken for different discharge currents, as shown in Fig. 4(b).

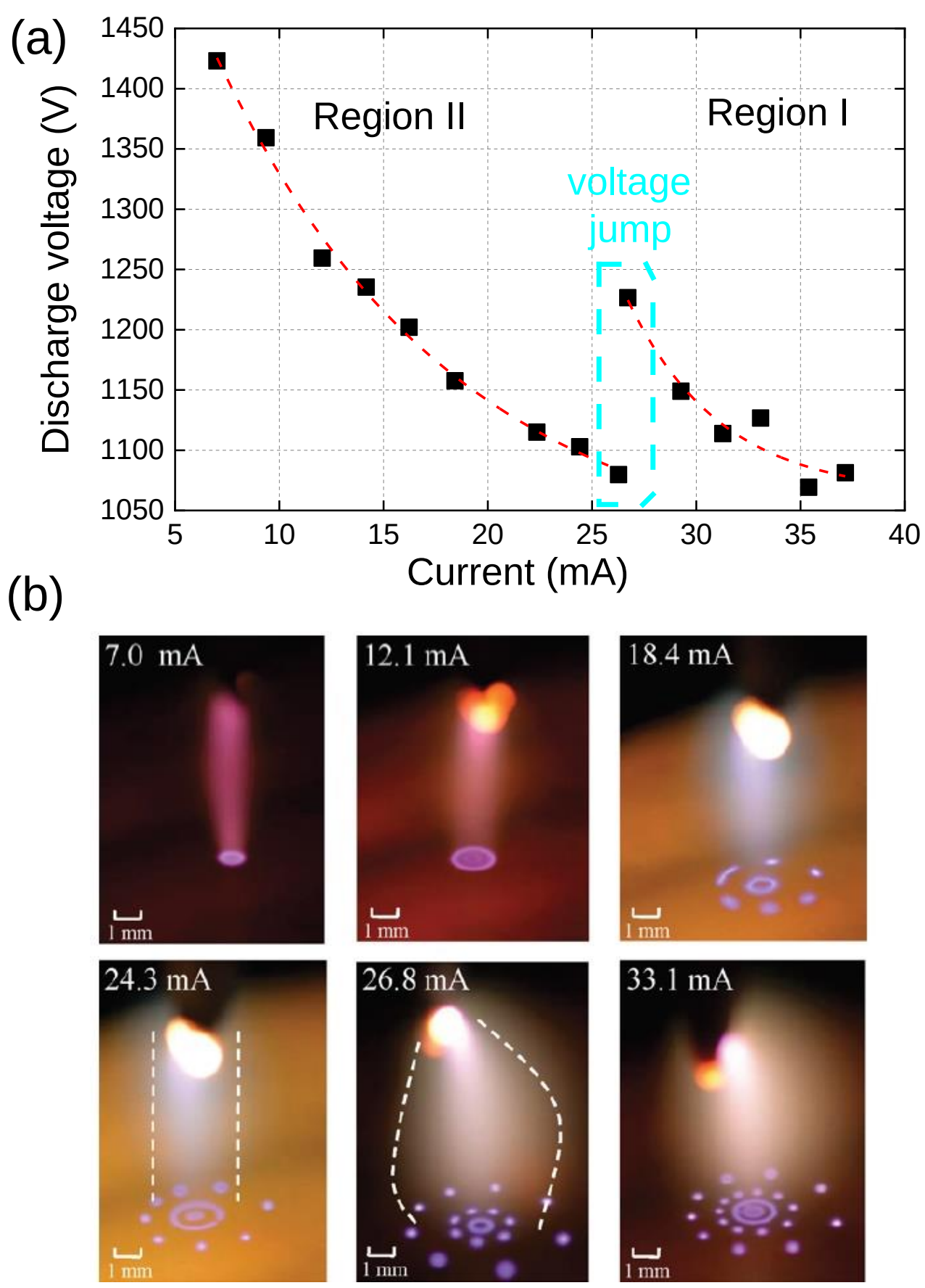

***Fig. 4. (a) V-I characteristic of the DC discharge and (b) the pictures of the corresponding SOP obtained for several discharge currents. The gap distance between the pin and the liquid is 5 mm. Tap water is employed, and the electric conductivity is 2.36 mS/cm. The exposure time of the picture in (b) is 8 ms. The white dashed lines depict the discharge shape in two regions.***

### III.B.2. Discussions on SOPs evolution with discharge current

These SOPs can be described according to the parameters used in the studies reported in Table I, i.e., their size, their structural complexity (e.g., concentric rings,[19,34,36,37] strips,[34,37] spots,[19,34,37] or round area[19,34,35,37]), and their dynamics, e.g., rotations.[19,34,36] In region I (I>26 mA), the bulk discharge appears diffusive in a pear-shape[48,69] with an enlarged diameter close to the one of the SOPs formed on the liquid interface. This SOP shows a high complexity degree for larger currents. For example, at 33.4 mA, the pattern is structured onto 4 virtual concentrated circles: On the two outer circles, the plasma is discreet since it appears as spots distributed at equal distances from one another, while on the two inner circles, the plasma spots seem to recover, hence forming two uniform and distinct rings. Decreasing the current from 33.4mA to 26.8mA induces a loss in SOP complexity with a single outer discreet ring and two inner uniform rings. This uniform aspect could be correlated with the exposure time of the camera, which is as high as 8 ms. The dynamics of the inner circles could not only be faster than those of the outer circles but also scale on times much faster than 8ms. For this reason, one could consider this uniformity as solely apparent.

Transition from region I to region II is marked by a morphological change in the bulk plasma, which is now columnar close to the filamentary mode due to the larger electric field (E/N). The columnar discharge can be seen clearly in Figs. 4(b) at 12.1mA and 7.0mA. At discharge currents 18.4 and 24.3mA, the discharge is close to the filamentary discharge. In this region II, however, SOPs are still present with a lower pattern diameter and lower structural complexity. For currents at 18.4 and 24.3mA, SOPs show distinct and concentric rings although much lower in number and complexity. For currents lower than 12mA, the SOP is no more designed as concentric rings but as a unique and uniform spot whose diameter is close to the one of the bulk discharges, namely, 1.2mm.

SOPs at the interface evolve continuously, while the discharge current increases without presenting two distinct regions. Therefore, the discharge morphology change seems not directly correlated with SOP size and discretization. The decrease in the SOP diameter seems correlated with the decrease in the discharge current and therefore with a decrease in the electron density in the bulk discharge as discussed in Sec. IIIC. From the point of ionization waves, if larger discharge currents lead to stronger ionization waves, more outer spots of the SOP would have a stronger electric field, enabling there a higher heating of electrons and enhancing the emission of these outer spots. However, this relationship between discharge current and ionization waves still needs further validation.

## III. C. Effect of the gap distance on SOP

### III.C.1. Discharge characteristics and SOP appearance

As shown in Fig. 5(a), increasing the gap distance from 1 to 11mm induces a linear decrease in the discharge current and a rise in the plasma power. Figure 5(b) shows the pictures of SOPs taken when increasing the gap distance. For the extreme gap distance (i.e., 1 and 10 mm), SOPs show diameters lower than 2mm with a poor discretization level. At intermediate gap distances such as 5–6 mm, SOP diameters become closer to 4mm and present a higher level of discretization.

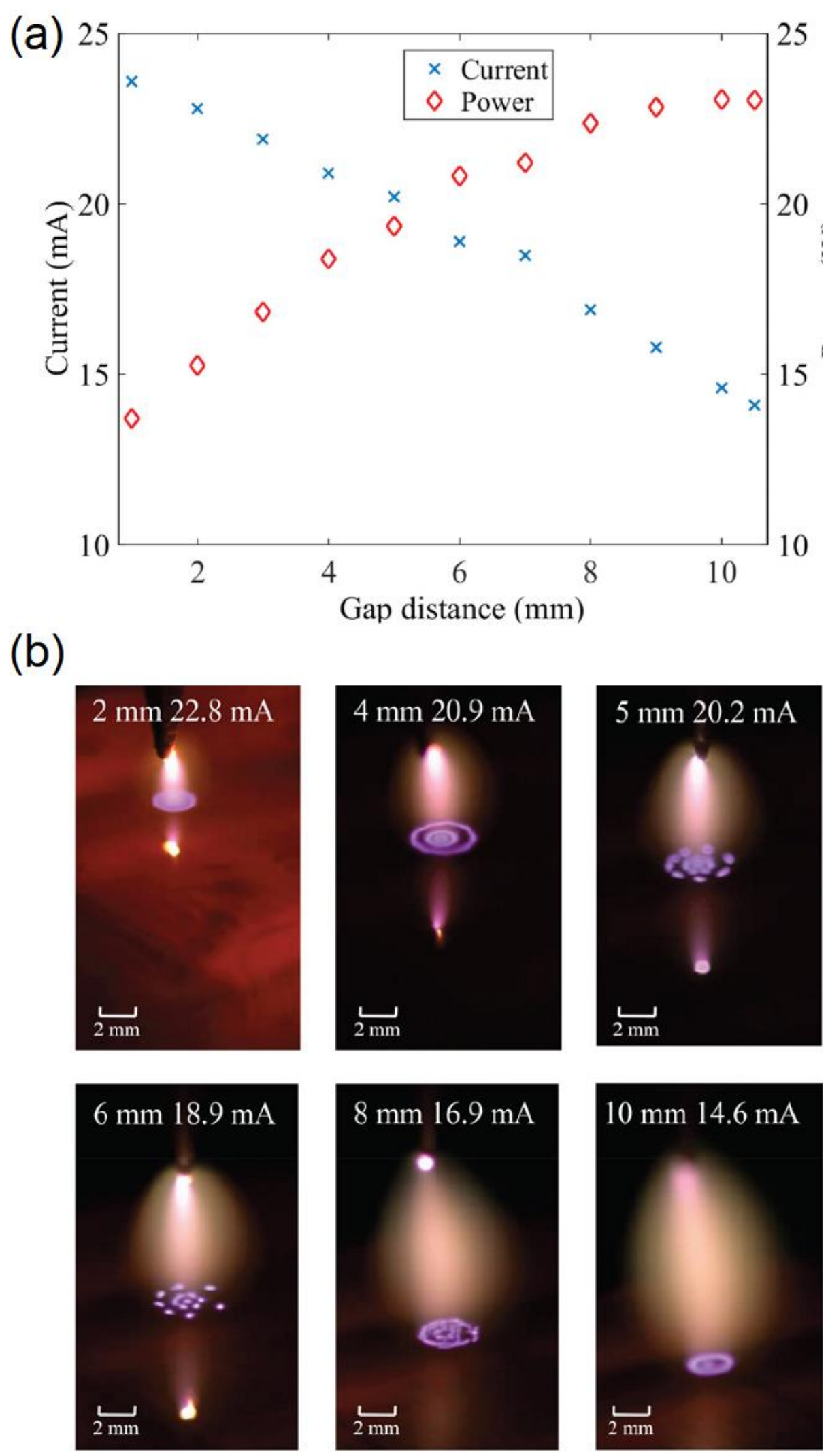

*Fig. 5. (a) Current and power dependence on the gap distance and (b) the corresponding formed SOP. The overall discharge voltage is 3200 V. Tap water is employed, and the electric conductivity is 3.08 mS/cm. The exposure time of the images in (b) is 8 ms, and the gap distance is also marked on each image.*

### III.C.2. Discussion on the SOPs' evolution with the gap distance

To sustain the DC discharge at larger gap, a higher discharge voltage would be necessary. However, the average electric field in the plasma bulk would simultaneously decrease, hence leading to a plausible decrease in the discharge current according to (1):

$$j = e.n_e.\mu_e.E \quad (1)$$

Changing the gap distance can simultaneously influence the two features of SOPs: SOP diameter and electron density. Their correlation can be studied by crudely estimating the electron density based on Eq. (1) with the following assumptions:

(i) The current density is determined considering the current values reported in Fig. 5(a) and the area of the SOP at the liquid interface.

(ii) Electric field intensity can reasonably be considered as equivalent to that in the plasma bulk since the local increase in the electric field close to the liquid anode takes place within 0.1mm (see above[46])

(iii) The electron mobility is referred to Ref. 49.

(iv) It should be noted that the obtained electron density is equivalent as it cannot discern the negative particle from the electron itself, and here, it is noted as $\bar{n}_e$.

Fig. 6 reports a clear correlation between SOP diameters, $\bar{n}_e$ estimations, and different gap distances. Larger SOP diameters are obtained for intermediate gap distances (3 to 7mm) and for minimum $\bar{n}_e$, around $10^{17}$ m$^{-3}$. The smallest diameters of SOPs are obtained at the extreme gap distance (1 and 10.5 mm) and the highest $\bar{n}_e$ value, ≈$3\times10^{17}$ m$^{-3}$. This correlation could also be validated by the results in Sec. III B (not shown in the work).

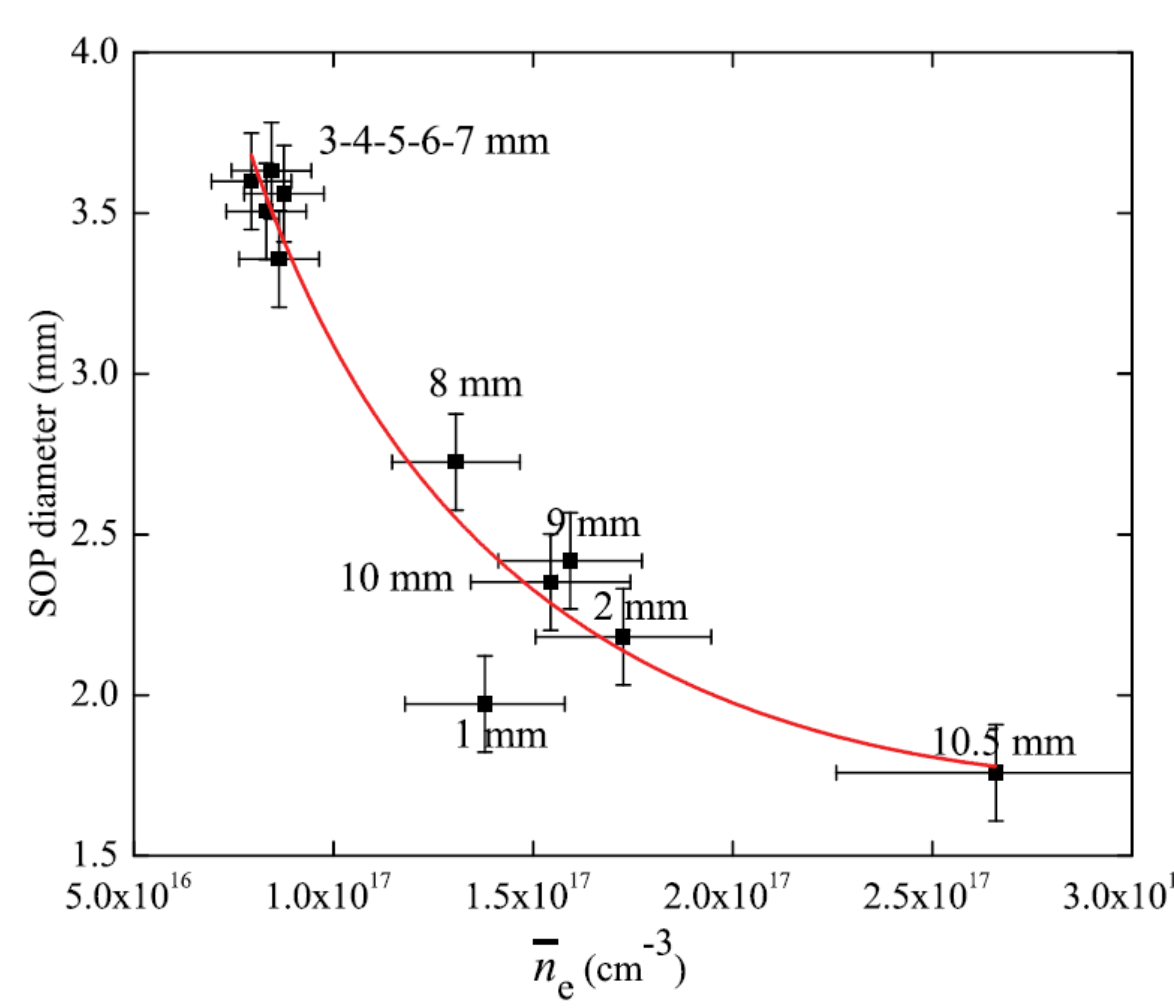

*Fig. 6. SOP diameter vs estimated electron density for several gap distances. The gap distance is marked for each data point. The fitting curve is s=1/(0.2464.log($\bar{n}_e$)– 9.322) with 95% confidence bounds.*

## III.D. Effect of the liquid type on SOP

To verify the SOP dependence on liquid types, a series of experiments are performed where the discharge and setup parameters are identical (same current, same gap distance, and same pin electrode) except the three liquids investigated: sulfuric acid, hydrochloric acid, and hydrofluoric acid. Their pH and EC are listed in Table III.

| EC | pH values | |
|---|---|---|
| (mS/cm) | $H_2SO_4$ solution | HCl solution |
| 10 | 1.56 | 1.57 |
| 57 | 0.93 | 0.89 |
| 150 | 0.40 | 0.33 |

***Table III. EC and the corresponding pH of two types of liquids. Experiments on the other electrolyte solutions have also been performed using NaCl, NaOH, $(NH_4)_2SO_4$ and $H_2SO_4$ solutions although their effects on SOPs show no unambiguous difference under the same electrical conductivity***

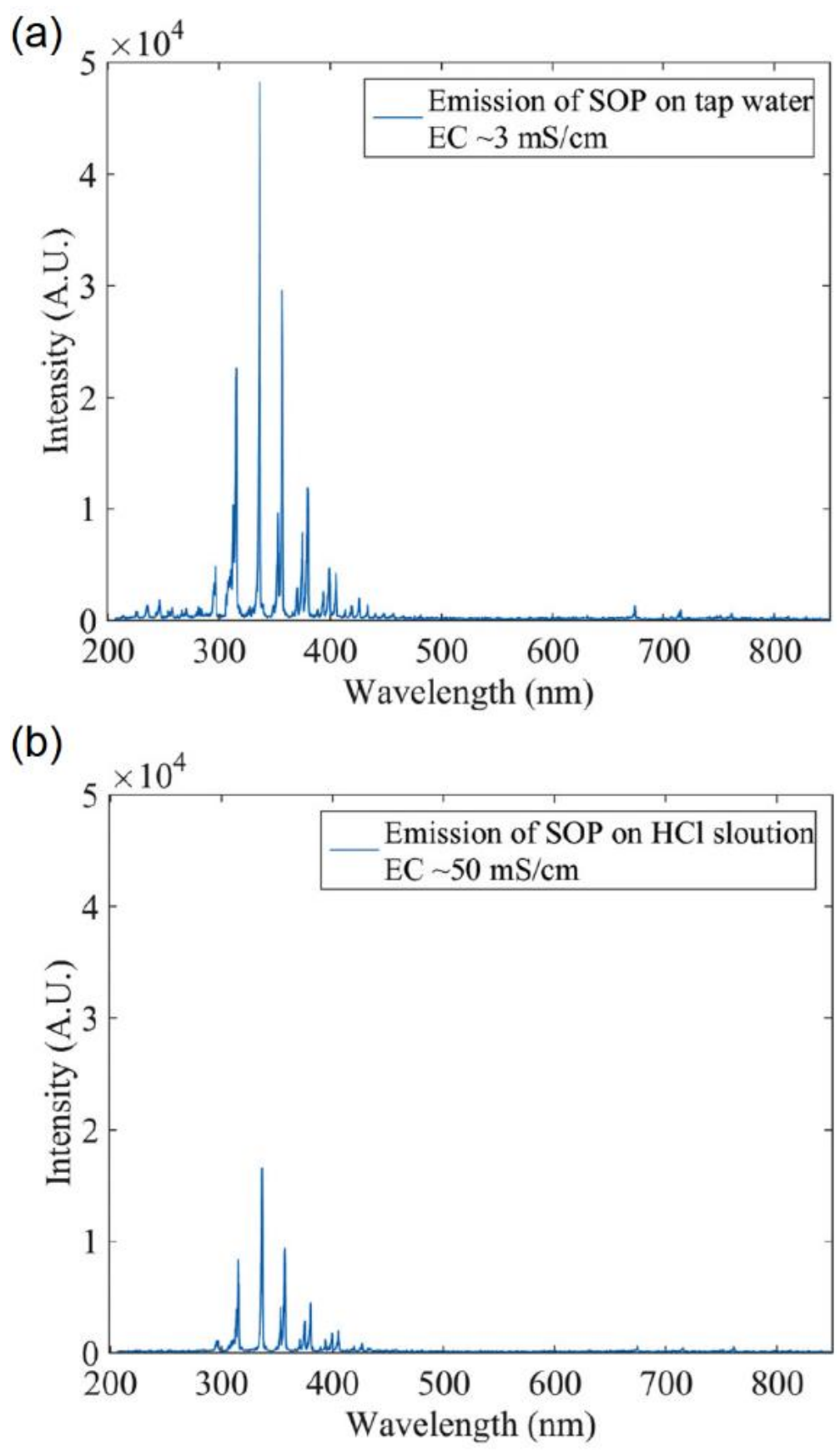

***Fig. 7. Emission spectra of SOP from (a) tap water and (b) HCl solution. The exposure time is 50 ms. Note that the emission spectra of SOP from $H_2SO_4$ solution are also the same as those from HCl solution with only minor different intensity.***

### III.D.1. SOPs emission spectra under different liquid types

As shown in Fig. 7, changing the liquid type (HCl with 50 mS/cm and tap water with 3 mS/cm, $H_2SO_4$ solution) does not have influence on the emission spectra of SOPs: the same bands and lines are observed although with different intensities.

### III.D.2. SOP characterization with discharge current under $H_2SO_4$ and HCl solutions

The V–I curves using two acidic solutions at 57 mS/cm and pH (≈0.9) are shown in Fig. 8(a). The two curves overlap and show a decrease in the voltage with the current. Meanwhile, the corresponding SOP formation is shown in Figs. 8(b) and 8(c). Clearly, even with similar electrical properties of the bulk discharge and the EC, the pH values of the anode liquid and the SOPs formed on the $H_2SO_4$ and HCl interface are significantly different. On one hand, the former remains like those obtained with water: as the current increases, SOPs become more discrete. On the other hand, the SOPs formed on the HCl interface show always a roundish and uniformly emissive pattern.

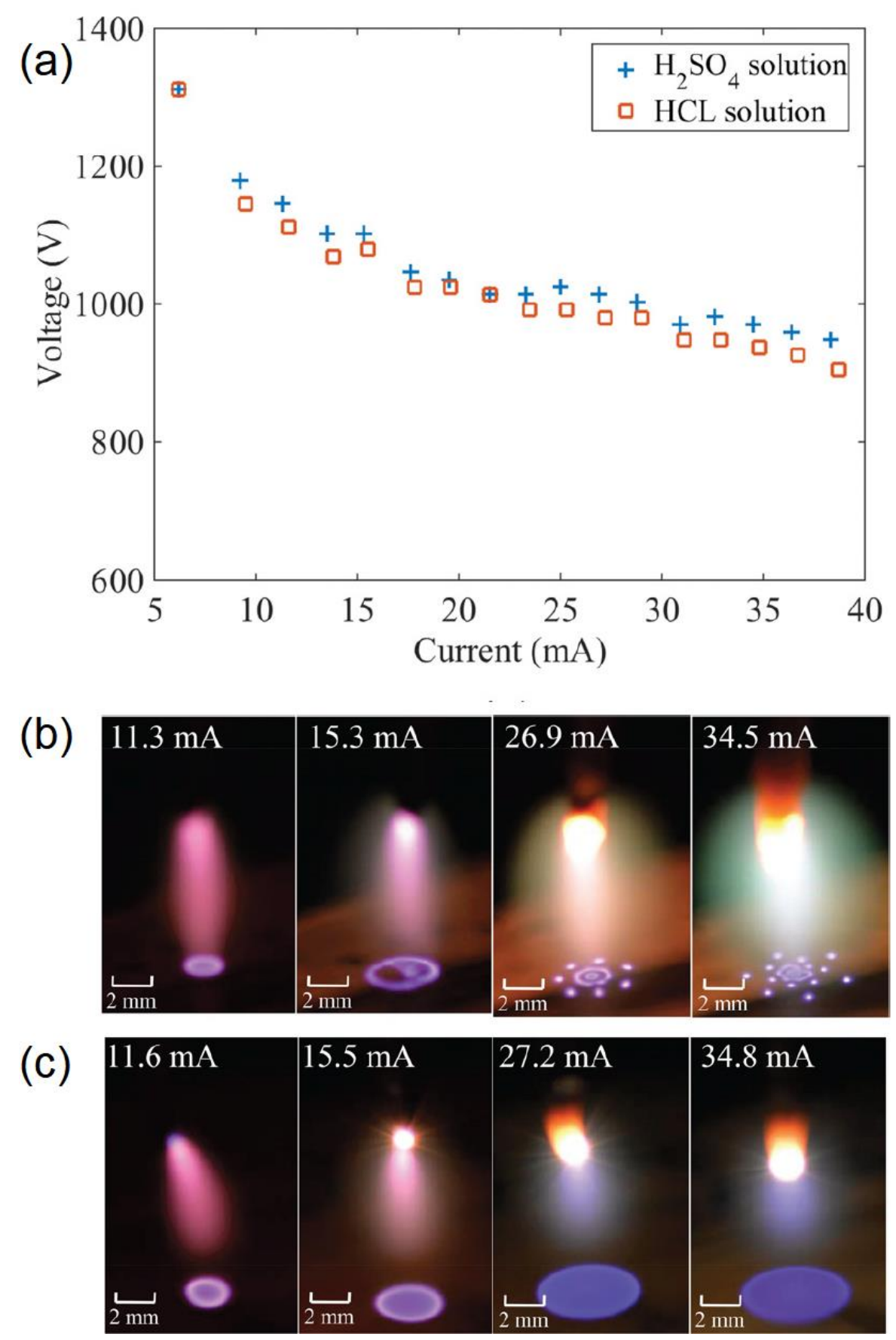

***Fig. 8. (a) V-I curve and pictures of SOPs generated on (b) $H_2SO_4$ and (c) HCl liquid surfaces, gap distance = 5mm, EC($H_2SO_4$) = EC(HCl) = 57 mS/cm, and pH($H_2SO_4$) = pH(HCl) ≈ 0.9***

#### III.D.3. SOP characterization with the gap distance under $H_2SO_4$ and HCl solutions

The effect of the gap distance on the discharge current and on SOPs is shown in Fig. 9, always considering $H_2SO_4$ and HCl solutions at the same electrical conductivity: 57 mS/cm. In Fig. 9(a), an overlap between the two curves is obtained between 1 and 12 mm. Consistent with Figs. 8(b) and 8(c), the SOP shows a discretization feature with $H_2SO_4$, while it appears as an apparent unique and large spot with HCl. Moreover, SOP diameters reach a maximum value for the intermediate gap distance for both liquids, i.e., 5–6 mm.

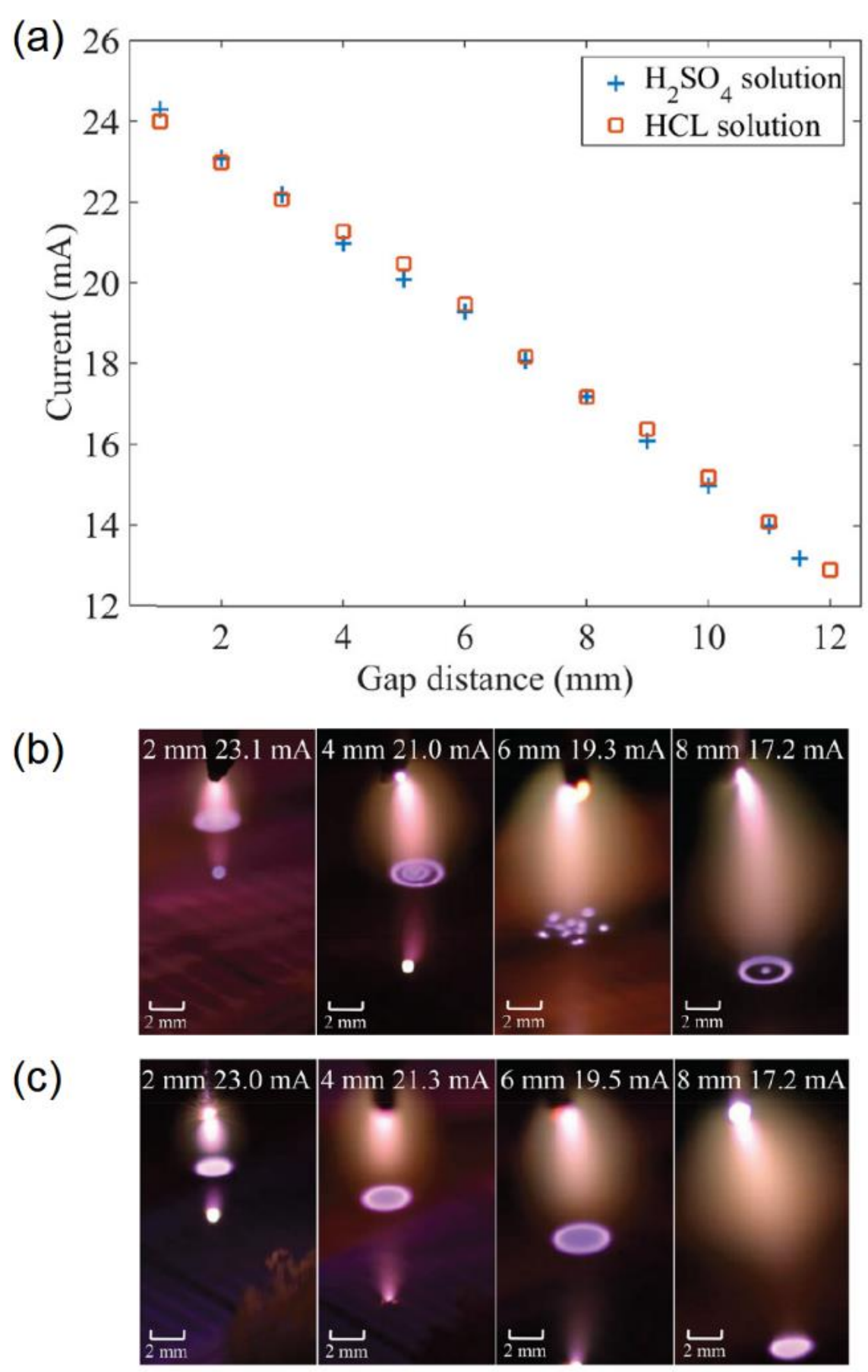

***Fig. 9. The dependence of (a) discharge current and SOPs generated with (b) $H_2SO_4$ and (c) HCl solutions on the gap distance. The discharge voltage is 3200 V, EC($H_2SO_4$) = EC(HCl) = 57 mS/cm, and pH($H_2SO_4$) = pH(HCl) ≈ 0.9.***

#### III.D.4. SOP characterization with $H_2SO_4$, HCl, and HF solutions

Figure 10 reports the formation of SOPs on $H_2SO_4$, HCl, and HF liquid interfaces at the same current (18.7mA 60.6 mA) and gap distances (6.3mm±1.0 mm) but for three electrical conductivities as follows:10, 57, and 150 mS/cm. The electrical conductivity of HF is not acquired with the HANNA conductivity probe to avoid its corrosion. It is estimated based on the HF concentration.[73]

The SOP difference between Figs. 8(b) and 8(c) and Figs. 9(b) and 9(c) still holds for greater EC such as 150 mS/cm with $H_2SO_4$ and HCl solutions. However, when the electric conductivity is lowered to 10 mS/cm, the SOP of HCl liquid shows a discretization feature as the same as the SOPs of $H_2SO_4$ solution as shown in Fig. 10(b). The additional SOP formation in Fig. 10(c) with HF solution on the other hand is more like SOP formation with HCl solution when the liquid conductivity increases from 10 mS/cm to 150 mS/cm.

#### III.D.5. Discussions on the SOP dependence on the liquid type

SOP patterns formed on a liquid surface can be characterized in terms of pattern diameter and discretization. According to SOP appearances in Figs. 8 and 9, the discharge current and gap distance can affect the SOP diameters greatly. The SOP size increases as the discharge current increases with a fixed gap distance. Moreover, the SOP size reaches the maximum at an intermediate gap distance. On the other hand, it is the SOP discretization that significantly depends on the liquid type. Meanwhile, the discretization feature determines the various types of SOPs, such as strips, rings, or spots.

As stated in the works of Shirai et al., replacing the ambient air by nitrogen makes the patterns (see Fig. 13 in Ref. 34) become similar to the ones we obtained when using HCl, as shown in Fig. 8(c). The explanation to their experiment relies on the attachment of free electrons. Since the gas contains no oxygen, the electron attachment strongly decreases to the benefit of higher density of free electrons, which can impinge on the liquid. Moreover, the discharge transits to a column shape if air is replaced by $N_2$. As a result, this discreet SOP is expected to be obtained only with smaller exposure time, not with 1/30 s in Ref. 34. In the case of the HCl liquid anode, the formed pattern is different from the $H_2SO_4$ liquid anode but with almost the same discharge current, shape, and gas environment. The underlying mechanism should be different.

Whatever the solution, one can clearly observe in Fig. 10 that (i) for EC as low as 10 mS/cm, the dark region appears narrow, while the SOP exhibits a discreet feature; (ii) for the two larger EC values, a thicker dark region is obtained, while the SOP shows a unique and large spot for HCl and HF solution but not for $H_2SO_4$; and (iii) whatever the EC or the acid considered, gas bubbles are generated from the immerged anode, evidencing potential electrolysis and therefore the decomposition of the liquid into reactive gaseous species, as reported in Table IV for water, $H_2SO_4$, HCl, and HF solutions.

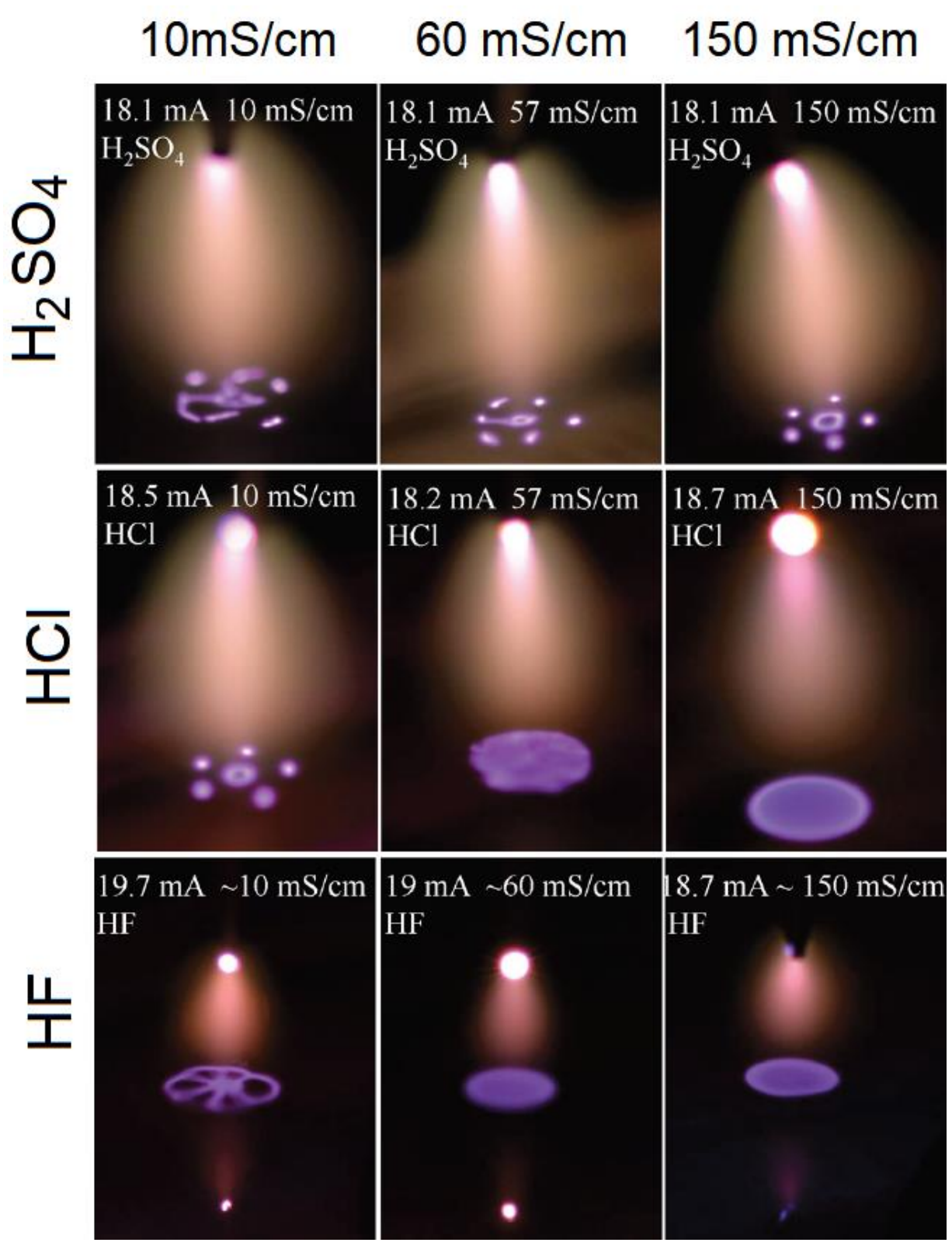

***Fig. 10. SOP dependence on the electric conductivity of (a) H2SO4, (b) HCl, and (c) HF electrolyte solutions. The gap distance for $H_2SO_4$ and HCl solutions is 7 mm, and the overall discharge is 3200 V. The gap distance for HF solution is 5 mm, and the discharge is 3200 V.***

As reported in Table IV, for tap water (low EC) and $H_2SO_4$ solution [high or low EC, also refer to Figs. 8(b) or 9(b)], $OH^-$ is expected to transfer electrons to the anode. For HCl and HF solution (high EC), $Cl^-$ and $F^-$ anions rather than $OH^-$ are oxidized (Table IV). With the highly concentrated HCl or HF electrolyte, gaseous $Cl_2$ and $F_2$ are generated and a low but non-negligible portion of the gases crosses the liquid interface into the ambient gas, which corresponds locally to the anode region of the DC discharge and the surrounding ambient air. Henri's law supports the fact that the thermodynamic disequilibrium at the interface is stronger in the case of $Cl_2$ and $F_2$ gases since ambient air does not contain any chlorine or fluorine species. The amounts of $Cl_2$ and $F_2$ diffused from liquid to gas may hence be slightly higher than $O_2$.[74]

| Electrolyte | Reactions on the anode |
|---|---|
| Tap water or $H_2SO_4$ solution | $4OH^-_{(liq)} \rightarrow 2H_2O_{(liq)} + O_{2(g)} + 4e^-$ |
| HCl solution | $2Cl^-_{(liq)} \rightarrow Cl_{2\ (g)} + 2e^-$ |

***Table IV. Reactions on the anode in different electrolytes***

Once these gases are released, they partly interact with the anode region of the discharge where the electron density is roughly estimated to $10^{17}$ cm$^{-3}$ (see Fig. 6). In this region, the electrons can interplay with the released electronegative gas, hence leading to the formation of negative ions. The electron affinities of $F_2$, $Cl_2$, and $O_2$ are 3.08 eV, 2.4 eV, and 0.45 eV, respectively.[75] Because $F_2$ and $Cl_2$ have higher electron affinities than $O_2$, one may consider that they penetrate further into the discharge than $O_2$. Larger electron affinities allow them to interact within the discharge on larger scales, not only on the anode region but also in the dark region, where they could make it thicker compared with tap water or even $H_2SO_4$ solution.

These negative ions resulting from electron affinity would move back to the liquid anode surface. The impact by these negative ions, which is also highly reactive with water itself, leads finally to different reaction on the water surface. Because SOPs are assumed to be a reaction-diffusion system,[43] different reactions would change SOPs. In a word, the electrolysis process in Table IV may lead to the different patterns of HCl and HF solutions.

As to the dependence on electric conductivity of the SOP, the local electric field at the interface might increase with the electrolyte concentration (electric conductivity here).[76] The increased electric field finally alters the behavior of the SOPs when EC of solution changes.

It should be noted that Table IV is obtained under the frame of conventional electrolysis. However, it should be kept in mind that the electro-chemical reactions of plasma in contact with liquid are different from those of the conventional electrolysis. For example, the electron temperature in this work is around 1 or 2 orders of magnitude higher than that of conventional electrolysis.[2] This high energy may result in efficient reaction $2H^+ + 2e^- \rightarrow H_2(g)$,[77] which should be the reason why there is no $H_\alpha$ or $H_\beta$ line observed at the interface of plasma and liquid in Sec. IIIA. Moreover, after electrons with several eV energy are solved into the interface, the hydrated electrons can involve more than one chemical reaction[2,78] and might lead to different gas production processes. Additionally, charge transfer via electronegative ions is also possible for plasma in contact with liquid. More difference could be referred to Refs. 2, 78, and 79.

# IV. Conclusions

The formation of SOPs in a DC pin liquid anode discharge in air was investigated. The discharge characteristics and the dependence of SOPs on discharge currents and the gap distance with the tap water anode were presented. A stratified discharge is formed in air with a dark region formed above luminous SOPs. Based on the emission spectra of the SOPs, it is the transition of $N_2(C^3\Pi - B^3\Pi)$ that forms the SOP emission. Meanwhile, $N_2(C)$ is excited in air by energetic electrons, which should be heated due to the local increase in the electric field at the interface.

The size of the SOPs would increase with the discharge current when the gap distance is fixed. If the gap distance increases, the size has the maximum area at the intermediate gap distance. The size evolution of SOPs quite reciprocally correlates with the local electron density as is simulated.

The SOPs behave discrete (more isolate spots) when the discharge current increases at a certain gap distance. However, when the gap distance changes, the SOPs have a higher discrete degree at an intermediate distance.

However, this discrete feature has a liquid type dependence. For HCl and HF solutions with an electric conductivity larger than 57 mS/cm, the SOPs hold a whole round area (large and single spot) with almost no visible discreteness. For $H_2SO_4$ solution, SOPs behave as tap water no matter how large the electric conductivity is. The relationship between the SOPs and electrolysis is initially discussed. This discovery may shed light on the correlation between the formation mechanisms of SOPs and the electrolysis process in the solution.

# V. Acknowledgments

This work was carried out within the LABEX Plas@Par project and received financial state aid managed by the Agence Nationale de la Recherche, as part of the Programme Investissements d'Avenir (PIA) under the Reference No. ANR-11-IDEX-0004-02. The authors thank F. Leblanc for providing the DC power supply. S. Z. thanks Professor Peter Bruggeman for the scientific advice. The authors also greatly thank the referees for their valuable comments.